\documentclass[doublecol]{epl2} 

\newcommand{\beq}{\begin{equation}}
\newcommand{\eeq}{\end{equation}}
\newcommand{\bsy}{\boldsymbol}

\title{Networks with arbitrary edge multiplicities}

\author{Vinko Zlati\'{c}\inst{1,2} \and Diego Garlaschelli\inst{3} \and Guido Caldarelli\inst{1,4,5}}
\shortauthor{V. Zlati\'{c} \etal}

\institute{                    
  \inst{1} ISC-CNR, Dipartimento di Fisica, Universit\`a ``La Sapienza'', P.le Moro 5, 00185 Roma, Italy\\
  \inst{2} Theoretical Physics Division, Rudjer Bo\v{s}kovi\'{c} Institute, P.O.Box 180, HR-10002 Zagreb, Croatia\\
  \inst{3} Lorentz Institute for Theoretical Physics, University of Leiden, Niels Bohrweg 2, NL-2333 CA Leiden,
 The Netherlands\\
  \inst{4} LINKALAB, Via San Benedetto 88, 09129 Cagliari, Italy\\
  \inst{5} London Institute for Mathematical Sciences, 22 South Audley Street, Mayfair, London W1K 2NY, United Kingdom
}

\abstract{
One of the main characteristics of real-world networks is their large clustering. 
Clustering is one aspect of a more general but much less studied structural organization of networks, i.e. edge multiplicity, defined as the number of triangles in which edges, rather than vertices, participate.  
Here we show that the multiplicity distribution of real networks is in many cases scale-free, and in general very broad.
Thus, besides the fact that in real networks the number of edges attached to vertices often has a scale-free distribution, we find that the number of triangles attached to edges can have a scale-free distribution as well. 
We show that current models, even when they generate clustered networks, systematically fail to reproduce the observed multiplicity distributions.
We therefore propose a generalized model that can reproduce networks with arbitrary distributions of vertex degrees and edge multiplicities, and study many of its properties analytically.}

\begin{document}

\maketitle

\section{Introduction}
Real networks, where nodes (or \emph{vertices}) are intricately connected by links (or \emph{edges}), are characterized by complex topological properties such as a scale-free distribution of the \emph{degree} (number of edges reaching a vertex), degree-degree correlations, and nonvanishing degree-dependent \emph{clustering} (density of triangles reaching a vertex) ~\cite{guidosbook}. 
Understanding the structural and dynamical properties of complex networks strongly relies on the possibility to investigate theoretical models which are both realistic and analytically solvable. 
Several analytically solvable models reproducing the most important local property of real networks, i.e. the degree distribution, have been proposed~\cite{guidosbook}. 
However, models reproducing higher-order properties including clustering (also called \emph{transitivity}) are only a few and are either entirely computational ~\cite{boguna0,Pusch} (i.e. not analytically solvable) or  solvable only for particular cases, e.g. when triangles are non-overlapping ~\cite{boguna1,boguna2,newman,adamic} or when the network is made by cliques ~\cite{gleeson} or other subgraphs ~\cite{karrer} embedded in a tree-like skeleton.
Unfortunately, real networks generally violate the above particular conditions, as empirical analyses have revealed and as we will further show in what follows.
Moreover, it has been shown that clustering is only one aspect of a more general
topological organization which is best captured by \emph{edge multiplicity}
~\cite{boguna1,boguna2,biasedWalks}, i.e. the number of triangles in which edges, rather
than vertices, participate. 
Besides being more informative than vertex-based clustering, edge multiplicity strongly determines the percolation properties of networks ~\cite{boguna3} and their community structure ~\cite{community}.

\section{A model with arbitrary edge multiplicities}
In order to overcome these limitations, here we propose an analytically solvable model of  networks with no restriction on their clustering properties, and able to generate  edges of any multiplicity. Let us denote by $m(i,j)$ the multiplicity of the edge $(i,j)$, i.e. $m(i,j)=\sum_{k\ne i,j}a_{ik}a_{kj}$ where $a_{ij}=1$ if a link between vertices $i$ and $j$ is there, and $a_{ij}=0$ otherwise. 
In our model we allow each vertex $i$ to have $k_i^{(0)}$ edges of zero multiplicity, $k_i^{(1)}$ edges of multiplicity $1$ and so on, up to $k_i^{(M)}$ edges of multiplicity $M$, where $M=N-2$ is the maximum possible multiplicity in a network with $N$ vertices.  
Thus each vertex $i$ is assigned a ($N-1$)-dimensional vector $\bsy{k}_i\equiv(k_i^{(0)},...,k_i^{(M)})$, that
we denote as the \emph{generalized degree}, specifying the multiplicity structure in the
neighborhood of $i$. The ordinary degree of vertex $i$ is $k_i=\sum_{m=0}^M k_i^{(m)}$.
Accordingly, we consider the ensemble of random networks with a specified distribution $P(\bsy{k})\equiv P(k^{(0)},k^{(1)},\dots,k^{(M)})$ of generalized degrees. 

\begin{figure}[t]
\onefigure[width=0.48\textwidth]{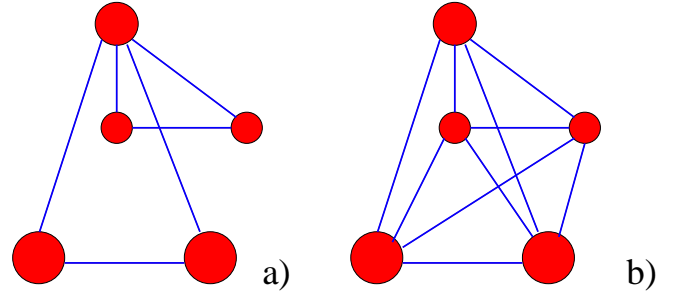}
\caption{\textbf{a)} Maximally clustered configuration ($c=1/3$) allowed for for the top vertex ($k=4$) in networks with non-overlapping triangles (weak transitivity) such as Newman's model ~\cite{newman}. 
\textbf{b)} Maximally clustered configuration ($c=1$) for the top vertex ($k=4$) in networks with overlapping triangles (strong transitivity), which is achieved in our model by assigning that vertex a generalized degree $\bsy{k}=(0,0,0,4,0,\dots)$.}
\label{Figure.1} 
\end{figure}

Our approach reduces to various previously proposed models in special cases, but 
is more general and allows to analytically investigate more realistic regimes which 
have not been explored so far. 
\begin{itemize}
\item
If $\bsy{k}_i=(k_i^{(0)},0,\dots,0)$, our model reduces to the \emph{configuration model} ~\cite{Molloy,Newman01} where each vertex $i$ has a specified degree $k_i=k_i^{(0)}$, and the network is locally tree-like (edges have zero multiplicity). This model has vanishing clustering in the thermodynamic limit $N\to\infty$, and is thus inadequate to reproduce clustered networks. 
\item
If $\bsy{k}_i=(k_i^{(0)},k_i^{(1)},0,\dots,0)$, Newman's clustered model ~\cite{newman} is recovered, where each vertex $i$ has attached $k_i^{(0)}=s_i$ ``single'' edges and $k_i^{(1)}=2t_i$ edges belonging to $t_i$ triangles. 
Although this model has a finite clustering for $N\to\infty$, it can only produce networks in the \emph{weak transitivity} regime ~\cite{boguna1,boguna2}, i.e. where the clustering coefficient of a vertex with degree $k$ is $c(k)\le (k-1)^{-1}$ (see figure ~\ref{Figure.1}a). 
\item
If $\bsy{k}_i=(0,k_i^{(1)},0,\dots,0)$, we recover the model by Shi et al. ~\cite{adamic} where all triangles are closed. This model is the maximally clustered version of Newman's model, i.e. $c(k)= (k-1)^{-1}$, but still cannot produce strong transitivity.
\item
If $\bsy{k}_i=(k_i^{(0)},0,\dots,0,k_i^{(c-2)},0,\dots,0)$ we recover Gleeson's model ~\cite{gleeson} where each vertex $i$ belongs to a clique of $c$ vertices (and thus has $k_i^{(c-2)}=c-1$ links of multiplicity $c-2$) and has $k_i^{(0)}=k_i-c+1$ additional external links of zero multiplicity, thus forming a network where cliques are embedded in a tree-like structure.
Although this model can produce networks with strong transitivity, it forces any vertex to  belong to only one clique. Thus it fails to reproduce networks with overlapping communities of densely interconnected vertices ~\cite{community}. 
\item
Finally, if $\bsy{k}_i=(k_i^{(0)},k_i^{(1)},k_i^{(2)},0\dots,0)$ we recover the model recently proposed by Karrer and Newman ~\cite{karrer} where, in addition to single edges and edges belonging to triangles, edges belonging to \emph{diamonds} (thus with multiplicity 2) are also introduced. More generally, that model allows to embed any type of small subgraphs into a higher-order tree-like structure, and can thus produce strong transitivity as in Gleeson's model. However, the model can only be applied as long as the set of specified subgraphs is fixed \emph{a priori}, and its analytical complexity grows rapidly with the number and size of the subgraphs considered. The empirical results that we will show in a moment make this approach inadequate to reproduce real networks.
\end{itemize}

\section{Edge multiplicity in real networks}
\begin{figure}[t]
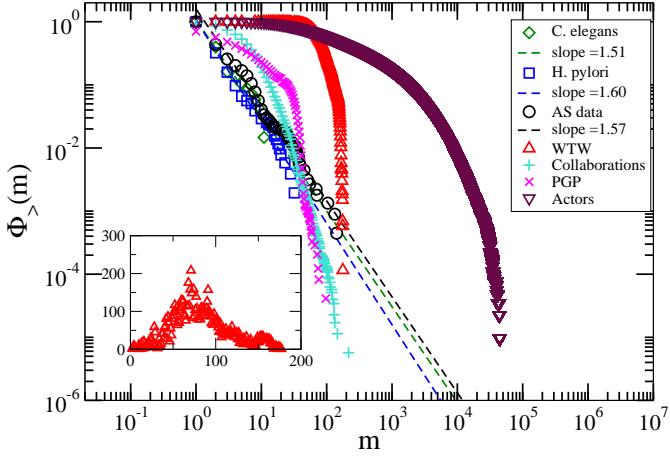

\onefigure[width=0.49\textwidth]{fig2.eps}
\caption{Cumulative edge multiplicity distributions $\Phi_>(m)$ for various real networks.
Inset: histogram of edge multiplicities (non-cumulative distribution) for the World Trade Web (WTW), as an example of network with unusually high density.}
\label{Figure.2} 
\end{figure}

In all the above models, the fraction $\Phi(m)$ of edges with multiplicity $m$ is fully concentrated on the smallest possible values, i.e. $m=0,1,2$ depending on the particular model (except in Gleeson's model, where a broader distribution of multiplicities can be obtained with a suitable choice of clique sizes, however losing an important degree of freedom required in order to fit other properties of real networks ~\cite{gleeson}). 
It is important to compare this prediction with the multiplicity structure of real networks.
In fig.\ref{Figure.2} we show the cumulative edge multiplicity distribution $\Phi_>(m)\equiv \sum_{n\ge m}\Phi(n)$ for various real networks. 
We find that sparse networks, such as the Internet and metabolic networks, display a power-law distribution of edge multiplicities (with similar exponents). 
Denser networks such as the World Trade Web show a distribution which is peaked at some very large value (see inset). 

In these and all other cases shown, the distributions are broad and extend over many orders of magnitude, in sharp contrast with the predictions of the above models. In particular, scale-free multiplicity distributions imply that, in models with modules embedded in tree-like structures, subgraphs of any size should be attached to vertices in order to reproduce the observed multiplicity structure. In this situation, such models become analytically intractable and their very philosophy becomes inappropriate. 
Indeed, the empirical results shown above suggest that network formation is much more decentralized than assumed by locally generating non-overlapping modules of fixed size and sparsely connecting them to one another. The concept of module itself appears vague, due to the lack (or to the unreasonable largeness) of a typical scale for the subgraphs required to describe the network.
Remarkably, besides the fact that in real networks the number of edges attached to vertices often has a scale-free distribution, we found that \emph{the number of triangles attached to edges can have a scale-free distribution as well}. 

\section{Generating functions and clustering}
Our model, by allowing $\bsy{k}$ to have a more general structure, can span the entire multiplicity spectrum without explicitly introducing subgraphs, overcoming the limitations of the aforementioned models (see figure ~\ref{Figure.1}b). 
The ordinary degree distribution is
\beq\label{eq. 1}
p(k)=\sum_{\bsy{k}}P(\bsy{k})\delta_{k,\sum_{m=0}^{M}k^{(m)}}
\eeq
The generating function of the probability $P(\bsy{k})$ is
\beq\label{eq. 2}
g(\bsy{z})=\sum_{\bsy{k}}(\bsy{z}\wedge\bsy{k})P(\bsy{k})
\eeq
where $\bsy{z}\wedge\bsy{k}=\prod_{m=0}^{M}z_m^{k^{(m)}}$ and $g(\bsy{z})=g(z_0,\dots,z_M)$. 
The generating function of the degree distribution is
\beq\label{eq. 3}
G(z)=\sum_{k=0}^{\infty}z^k\sum_{\bsy{k}} P(\bsy{k})\delta_{k,\sum_{m=0}^{M}k^{(m)}}=g(z,z,...,z)
\eeq
We can now compute the transitivity of the network.
First we need to count the triangles:
\beq\label{eq. 4}
3N_{\bigtriangleup}=N \sum_{m=0}^M \left[\frac{\partial g(\bsy{z})}
{\partial z_m}\right]_{\bsy{z}=1}\!\!\!\bsy{e}_{m}\cdot\bsy{I}=N\bsy{I}\cdot\nabla
g(\bsy{z})\mid_{\bsy{z}=\bsy{1}}
\eeq
where $\bsy{e}_{m}$ is a \emph{unit vector} of \emph{multiplicity} $m$ (i.e. $\bsy{e}_{0}\equiv(1,0,...,0)$, $\bsy{e}_{1}\equiv(0,1,0,...,0)$, etc.) and
$\bsy{I}=\sum_{m=0}^M m\bsy{e}_{m}$. The total number of connected triples is
\beq\label{eq. 5}
N_3=\frac{N}{2} \frac{\partial^2 G(1)}{\partial z^2}
\eeq
so that the transitivity, which is defined as
$T=3N_{\bigtriangleup}/N_3$, does not disappear when $N\rightarrow \infty$. 
Therefore, as expected, our model successfully produces networks with non-vanishing overall clustering.
It can also generate any desired clustering spectrum, i.e. the average clustering $\bar{c}(k)$ of vertices with degree $k$. The latter is
\begin{equation}\label{LocalClust}
\bar{c}(k)=\frac{1}{Np(k)}\sum_{i=1}^N\frac{2N_{\bigtriangleup}(i)}{k(k-1)}\delta_{k_i,k}
\end{equation}
where $N_{\bigtriangleup}(i)$ is the number of mutually connected neighbors of vertex $i$. 
This leads to 
\begin{equation}\label{ClustRel}
\frac{k(k-1)}{2}\bar{c}(k)p(k)=\sum_{\bsy{1}\cdot\bsy{k}=k}\bsy{I}\cdot\bsy{k}P(\bsy{k}). 
\end{equation}
The above relations hold for every network. It is thus possible to choose $P(\bsy{k})$ in order to reproduce both $p(k)$ and $\bar{c}(k)$ as in other models ~\cite{boguna0,Pusch,gleeson}.

\section{Percolation properties}
Importantly, we can study the percolation properties of our model analytically, thus extending previous results ~\cite{Pusch,newman,gleeson,boguna3} to more general cases.
Let $D(s|\bsy{k})$ be the probability that a vertex of generalized degree $\bsy{k}$ is a member of a set of $s$ mutually reachable vertices. 
Similarly, let $d(s|\bsy{k})$ be the probability that a vertex connected to a vertex $v$ of generalized degree $\bsy{k}$ can reach $s$ other vertices, excluding the vertex $v$ and its neighborhood. 
The relation between $D(s|\bsy{k})$ and $d(s|\bsy{k})$ is
\begin{equation}
\label{eq. 7}
D(s|\bsy{k})=\sum_{s_1,\dots,s_k}d(s_1|\bsy{k})\cdots d(s_k|\bsy{k})\delta_{s,1+s_1+\ldots+s_k}. 
\end{equation}
We can also write a recursion relation for $d(s|\bsy{k})$ as
\begin{eqnarray}\label{eq. 8}
d(s|\bsy{k})
&=&\sum_{\bsy{h}}\sum_{m=0}^{min(h,k)-1}p(\bsy{h},m|\bsy{k})\nonumber\\
&\times&\sum_{s_1,\dots,s_{h_r}}d(s_1|\bsy{h})\cdots d(s_{h_r}|\bsy{h})\delta_{s,1+s_1+\ldots+s_{h_r}}
\end{eqnarray}
where $p(\bsy{h},m|\bsy{k})$ represents the probability to select, around a vertex of generalized degree $\bsy{k}$, an edge of multiplicity $m$ leading to a vertex of generalized degree $\bsy{h}$. 
The reduced degree $h_r$ is the number of vertices attached to the destination vertex except itself and the neighborhood of the first vertex i.e. $h_r=h-m-1$.
If degree-degree correlations can be neglected, $p(\bsy{h},m|\bsy{k})$ reads 
\begin{equation}\label{eq. 9}
p(\bsy{h},m|\bsy{k})=\frac{k^{(m)}}{k}\frac{h^{(m)}P(\bsy{h})}{\langle k^{(m)}\rangle}.
\end{equation}
The first fraction in eq.(\ref{eq. 9}) represents the probability to leave 
a vertex of generalized degree $\bsy{k}$ following an edge of multiplicity $m$. The second fraction is the probability to reach a vertex of generalized degree $\bsy{h}$ following that edge.
We can also use eq.(\ref{eq. 8}) to write the 
generating functions $\hat{d}(z|\bsy{k})=\sum_sz^sd(s|\bsy{k})$ of the probabilities $d$:
\begin{equation}\label{eq. 10}
\hat{d}(z|\bsy{k})=z\sum_{\bsy{h}}\sum_{m=0}^{min(h,k)-1}p(\bsy{h},m|\bsy{k})\left[\hat{d}(z|\bsy{h})\right]^{\bsy{h_r}}.
\end{equation}
If eq. (\ref{eq. 10}) has a stable solution $\hat{d}(z=1|\bsy{k})\leq 1$ the network percolates. 
In order to study the stability of eq. (\ref{eq. 10}) around $z=1$ we can study a perturbative 
solution $\hat{d}(z=1|\bsy{k})\approx 1+\chi(\bsy{k})\epsilon$ in the limit $\epsilon \rightarrow 0$, which yields
\begin{eqnarray}\label{eq. 13}
\chi(\bsy{k})&=&
\sum_{\bsy{h}}\sum_{m=0}^{min(h,k)-1} p(\bsy{h},m|\bsy{k})(h-m-1)\chi(\bsy{h})\\
&=&\sum_{\bsy{h}}\sum_{m=0}^{min(h,k)-1} \frac{k^{(m)}h^{(m)}}{\langle k^{(m)} \rangle k}(h-m-1)P(\bsy{h})\chi(\bsy{h})\nonumber\\
&=&\sum_{\bsy{h}}\left[\bsy{\alpha}\cdot\bsy{\beta}(h-1)-\bsy{\alpha}\cdot(\bsy{I}*\bsy{\beta})\right]P(\bsy{h})\chi(\bsy{h})\nonumber
\end{eqnarray}
where $\bsy{\alpha}=\bsy{k}/k$, $\bsy{\beta}=\sum_m\frac{h^{(m)}}{\langle k^{(m)}\rangle}\bsy{e}_m$ and 
$\bsy{I}*\bsy{\beta}\equiv\sum_mI_m\beta_m\bsy{e_m}$ is a vector.
The percolation transition occurs when the maximum eigenvalue of the matrix in eq.(\ref{eq. 13}) is larger than $1$ i.e $\Lambda_{max}>1$. 
Thus we have obtained an analytical expression for the percolation transition, more general than the one known for networks in the weak transitivity regime ~\cite{boguna2}, and valid for any level of clustering and multiplicity.

\section{Rich-club effect}
As another example of the effects of broad edge multiplicities, we now consider the
\emph{rich-club coefficient} $R(k)$, defined as the observed number of edges $E_>(k)$
among the $N_>(k)=Np_>(k)$ vertices with degree larger than $k$ (where $p_>(k)$ is the
cumulative degree distribution), divided by the maximum allowed number 
$N_>(k)(N_>(k)-1)/2\approx N^2p_>^2(k)/2$ ~\cite{Zhou,Colizza,myrichclub}.
In random networks with given degree distribution, the rich club behaves approximately as $R(k)_{Rand}\sim\frac{k^2}{\langle k \rangle N}$ ~\cite{Colizza}, so that the measured $R(k)$ must be compared to this non-constant value. 
We now consider the case when, as in our model, one also specifies a multiplicity
distribution $\Phi(m)$. Since every edge $(i,j)$ with multiplicity $m(i,j)\ge k$ surely
connects two vertices $i,j$ with degrees $k_i,k_j>k$, the expected value of $E_>(k)$ now
receives a contribution $E\Phi_>(k)$ from edges with multiplicity $m\ge k$ (where $E$ is
the total number of edges), and the standard approximation can only be applied to the
remaining $E(1-\Phi_>(k))$ edges. Following ~\cite{Colizza}, we obtain the modified
expectation
\begin{equation}\label{eq. 19}
R(k)_{Rand}\sim\Phi_>(k)\frac{\langle k \rangle}{Np_>^2(k)}+[1-\Phi_>(k)]\frac{k^2}{\langle k \rangle N} 
\end{equation}
If, as in some of the networks considered above, the cumulative distributions $\Phi_>(k)$ and $p_>(k)$ are power laws with exponents $-\alpha$ and $-\gamma$ respectively, the asymptotic behavior of the first summand is $\sim k^{2\gamma-\alpha}$ thus reducing or increasing the predicted scaling $\sim k^2$.

\section{Graphic generalized degree sequences}
There have been many attempts in the literature to generate null models of real networks by generating ensembles
of random graphs with given properties. Some of these approaches
make use of generating functions~\cite{Newman01,boguna1,boguna2}, as in the present paper.
Other approaches aim at constructing randomized ensembles computationally,
and generate so-called  \emph{microcanonical ensembles}~\cite{Maslov, Catanzaro, Gorka} of networks with sharp constraints. 
Finally, other approaches aim at describing random networks with given properties analytically, and generate \emph{(grand)canonical ensembles} of networks with soft constraints~\cite{Park,Diego1,Fronczak,ginestra,Diego2,Diego3}.

If our model is used as a null model for a particular real network, it gives predictions about the ensemble of random graphs having the same generalized degree sequence $\{\bsy{k}_i\}_{i=1}^N$ as the real network.
This generalizes the configuration model \cite{Molloy,Newman01} where only the ordinary degree sequence $\{k_i\}_{i=1}^N$ is specified. In the latter case, if $\{k_i\}_{i=1}^N$ is taken from a real network, one is sure that it is a graphic sequence. Otherwise, if one generates it artificially, one must enforce specific conditions, given by the Erd\H{o}s-Gallai ~\cite{erdos} and Havel-Hakimi ~\cite{hakimi} theorems, ensuring that the sequence is graphic. 
In our case, the realizability of $\{\bsy{k}_i\}_{i=1}^N$ is much more complicated than in the case of ordinary graphic degree sequences, but we now show how it can be related to the standard problem. For convenience, we define the $N\times (N-1)$ matrix $\mathbf{Q}$ with entries $Q_{ij}\equiv k_i^{(j-1)}$.
The row and column sums (i.e. the \emph{margins}) of $\mathbf{Q}$ are the degree sequence and the (unnormalized) multiplicity distribution respectively: \begin{eqnarray}
Q_{i+}&\equiv&\sum_{j=1}^{N-1}Q_{ij}=\sum_{m=0}^{N-2}k^{(m)}_i=k_i\\
Q_{+j}&\equiv&\sum_{i=1}^{N}Q_{ij}=\sum_{i=1}^{N}k^{(j-1)}_i=2E^{(j-1)}
\end{eqnarray}
where $E^{(m)}$ denotes the total number of edges with multiplicity $m$. 
Therefore, as a first condition we find that the marginal (ordinary) degree sequence $\{k_i\}_{i=1}^N$ must be graphic.
There are however strong additional constraints.
First note that, since we can always partition the edges in disjoint sets (each with given multiplicity), each of the $M$ sequences $\{k^{(m)}_i\}_{i=1}^N$ must be separately graphic. This introduces constraints along each column of $\mathbf{Q}$. 
Moreover, since edge multiplicities must be consistent with each other, there are also constraints along each row of $\mathbf{Q}$. 

A useful mapping allows us to solve the problem. For a given vertex $i$, we consider the subgraph $\Gamma_i$ whose vertices are the neighbors of $i$ and edges are their mutual connections. An example is shown in Fig.\ref{Figure.2} (note that $\Gamma_i$ does not contain  vertex $i$ itself). If we denote by $[x]_i$ the value of a topological property $x$ (e.g. the number $E$ of edges, or the link density $D=2E/[N(N-1)]$) when measured on the subgraph $\Gamma_i$, we find important relations, e.g.
\begin{equation}
[N]_i=k_i;\qquad [D]_i=c_i;\qquad [k_j]_i=m(i,j).
\end{equation}
\begin{figure}[t]
\onefigure[width=0.49\textwidth]{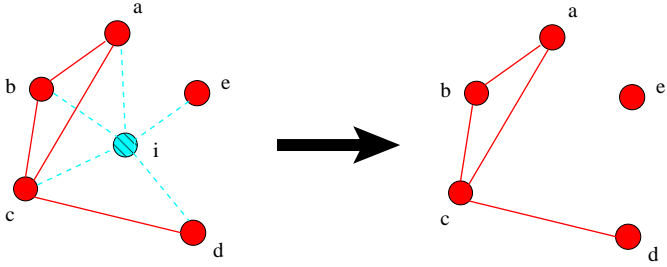}\caption{On the left side, a generic network with $N=6$ vertices is shown, and the edges attached to vertex $i$ are highlighted (cyan dashed edges).
The $(N-1)$-dimensional generalized degree of vertex $i$ is in this case $\bsy{k}_i=(1,1,2,1,0)$ (since the multiplicities of the dashed edges are $m_{ia}=2$, $m_{ib}=2$, $m_{ic}=3$, $m_{id}=1$, $m_{ie}=0$, and there is no edge with maximum multiplicity $N-2=4$).
On the right side, the $i$-associated subgraph $\Gamma_i$ is shown. The degree of each vertex $j$ in $\Gamma_i$ coincides with the multiplicity $m(i,j)$ of the edge connecting $j$ to $i$ in the original graph on the left.}
\label{Figure.3} 
\end{figure}

In other words, the number of vertices and link density of $\Gamma_i$ coincide with the degree and clustering coefficient of vertex $i$ measured on the whole network respectively.
Similarly, the degree of vertex $j$ in $\Gamma_i$ coincide with the multiplicity $m(i,j)$ in the whole network. 
Since there are $k^{(m)}_i$ vertices in $\Gamma_i$ whose degree $[k_j]_i$ equals $m$, $k^{(m)}_i$ is the unnormalized degree distribution of $\Gamma_i$, and the associated degree sequence $\{[k_j]_i\}_i$=$\{m(i,j)\}_i$ must therefore be graphic.
This observation enforces the required constraints along the rows of $\mathbf{Q}$  (and also shows that our model, by specifying the entire degree distribution of $\Gamma_i$, is a sort of configuration model for each graph $\Gamma_i$; by contrast, models that specify the clustering coefficient $c_i$ alone are a sort of Erd\H{o}s-R\'enyi random graph reproducing only the link density of $\Gamma_i$). 
Taking the two conditions together, we find that a necessary condition for a generalized degree sequence $\{\bsy{k}_i\}_{i=1}^N$ to be graphic is that, for fixed $m$, $k_i^{(m)}$ is a graphic degree sequence and, for fixed $i$, $k_i^{(m)}$ is a graphic degree distribution. 
This \emph{Sudoku}-like condition operates along each row and column of $\mathbf{Q}$
simultaneously.

\section{Conclusions}
In this paper we have shown that real networks display broad, and often scale-free, edge multiplicity distributions.
Existing models cannot reproduce such feature and are therefore inadequate to predict various properties of real networks. 
We have therefore introduced a model for networks with arbitrary generalized degree sequences. Unlike previous approaches, our model can take as input detailed information about the observed multiplicity structure to give refined analytical predictions about various network properties. 
We have finally exploited a useful mapping to give necessary conditions for a generalized degree sequence to be graphic.


\acknowledgments
This work was supported by FET Open project FOC nr. 255987. 
V.Z. also acknowledges support from the Croatian Ministry of Science, Education and Sports project nr. 098-0352828-2863.
D.G. also acknowledges support from the Dutch Econophysics Foundation (Stichting Econophysics, Leiden, Netherlands) with funds from beneficiaries of Duyfken Trading Knowledge BV, Amsterdam, Netherlands.

\end{document}